\documentclass[11pt,twoside]{article}   

\usepackage{setspace}
\usepackage[super,sort,comma,compress]{natbib}
\usepackage{siunitx}

\usepackage{fancyhdr}		

\usepackage[section]{placeins}   %
\usepackage{amssymb} 
\usepackage{graphicx}
\usepackage{amsmath}
\usepackage[margin=1.5in]{geometry}
\makeatletter \renewcommand\@biblabel[1]{$^{#1}$} \makeatother
\newcommand{\note}[1]{\mbox{}\\ \noindent \rule{16cm}{0.5mm} \\
{\em #1} \\ \noindent \rule{16cm}{0.5mm}
\typeout{    }
\typeout{***********note active on this page *************************}
\typeout{Note: #1  }
\typeout{****************************************end Note}
}

\newcommand{\cen}[1]{\begin{center} #1 \end{center}}

\usepackage[mathlines]{lineno}

\usepackage{hyperref}
\hypersetup{ colorlinks,
    citecolor=blue,
    filecolor=blue,
    linkcolor=blue,
    urlcolor=blue
}

\usepackage{xcolor}

\definecolor{gray}{rgb}{0.6,0.6,0.6}
\definecolor{red}{rgb}{0.85,0,0}
\definecolor{green}{rgb}{0,0.85,0}
\definecolor{blue}{rgb}{0,0,0.75}
\definecolor{beige}{rgb}{0.92,0.87,0.78}
\usepackage[all]{hypcap}    

\begin{document}
\cen{\sf {\Large {\bfseries Look-Up Table-Correction for Beam Hardening-Induced Signal of Clinical Dark-Field Chest Radiographs}}\\  
\vspace*{10mm}
Lochschmidt, Maximilian E.$^{1,2,3,*}$; Urban, Theresa$^{1,2,3}$; Kaster, Lennard $^{1,2,3}$; Schick, Rafael $^{1,2,3}$; Koehler, Thomas $^{4,5}$; Pfeiffer, Daniela$^{3,4}$ and Pfeiffer, Franz$^{1,2,3,4}$} 
(1) Chair of Biomedical Physics, Department of Physics, TUM School of Natural Sciences, Technical University of Munich, 85748 Garching, Germany;\\
(2) Munich Institute of Biomedical Engineering, Technical University of Munich, 85748 Garching, Germany; \\
(3) Institute for Diagnostic and Interventional Radiology, School of Medicine and Health, TUM Klinikum, Technical University of Munich (TUM), 81675 München, Germany; \\
(4) TUM Institute for Advanced Study, Technical University of Munich, 85748 Garching, Germany\\
(5) Philips Innovative Technologies, 22335 Hamburg, Germany
\vspace{5mm}\\

\pagenumbering{roman}
\setcounter{page}{1}
\pagestyle{plain}
* maximilian.lochschmidt@tum.de (M.E. Lochschmidt)\\
\begin{abstract}
\noindent {\bf Background:} 
The microstructure of material at a $\SI{}{\micro\meter}$ length scale leads to ultra-small-angle scattering of X-rays, which typically occurs, e.g., for lung tissue or some plastic foams. When using an interferometer, this effect alters the visibility of the fringe pattern, which can be detected and resolved by the detector. Thus, the ultra-small-angle scattering can be represented as a dark field image. For a polychromatic source, the hardening of the source spectrum changes visibility as well, generating an additional fake dark-field signal by the attenuation of the material on top of the real ultra-small-angle scatter-related dark-field signal.
Consequently, even homogeneous materials without microstructure typically exhibit a change in visibility.\\ 
{\bf Purpose:} The objective of this study is to develop a fast, simple, and robust method to correct dark-field signals and bony structures present due to beam hardening on dark-field chest radiographs of study participants.\\
{\bf Methods:} The method is based on calibration measurements and image processing. Beam hardening by bones and soft tissue is modeled by aluminum and water, respectively, which have no microstructure and thus only generate an artificial dark-field signal. Look-up tables were then created for both. By using a weighted mean of these, forming a single look-up table, and using the attenuation images, the artificial dark-field signal and thus the bone structures present are reduced for study participants.\\
{\bf Results:}  It was found that applying a correction using a weighted look-up table leads to a significant reduction of bone structures in the dark-field image. The weighting of the aluminum component has a substantial impact on the degree to which bone structures remain visible in the dark-field image. Furthermore, a large negative bias in the dark-field image—dependent on the aluminum weighting—was successfully corrected.\\
{\bf Conclusions:} Beam hardening–induced signal in the dark-field images was successfully reduced using the method described. The choice of aluminum weighting to suppress rib structures, as well as the selection of bias correction, should be evaluated based on the specific clinical question.\\

\end{abstract}
\note{Keywords: X-ray Imaging, Chest Radiography, Dark-field Imaging, Grating Based Imaging, Medical Physics, Medical Imaging}
\newpage     



\newpage

\pagenumbering{arabic}
\setcounter{page}{1}
\pagestyle{fancy}

\setlength{\headheight}{13.59999pt}
\addtolength{\topmargin}{-1.59999pt}
\section{Background}
When an X-ray beam with an energy used in medical imaging passes through matter, several physical interactions can change the beam's properties: Variations in the speed of electromagnetic radiation within the material cause a change in the beam direction, and photoelectric absorption attenuates the beam. Compton scattering is typically also considered a beam-attenuating effect. In contrast, ultra-small-angle x-ray scattering (USAXS), where the typical scatter angles are so small that scattered photons cannot be distinguished with conventional detectors from non-scattered photons, is most often considered separately. The scattering power of USAXS is influenced by the size and shape of the material's microstructure\cite{Feigin1987,Yashiro2010,Lynch2011,Strobl2014,Prade2015,Gkoumas2016,Graetz2020}. Phase shift and USAXS cannot be directly resolved by detectors with a typical pixel size of $\SI{150}{\micro\meter}$. However, if precisely arranged gratings are placed in the beam path, all three phenomena can be determined \cite{David2002,Momose2003,Weitkamp2005}.\\
It is important to note that the term "dark-field signal" or "dark-field image" typically encompasses all effects that change the visibility, such as the USAXS, beam hardening (BH), and Compton scattering. The objective is to ensure that the dark-field image exclusively conveys the USAXS information.\\
Although USAXS and attenuation are two distinct and independent phenomena, in the case of a polychromatic X-ray beam, attenuation can influence the USAXS channel by hardening the beam on attenuating materials\cite{DeMarco2024, Yashiro2015}. In the past, methods have been developed to correct the beam-hardening induced signal in phase-contrast and dark-field imaging using simulations and models for various single materials \cite{Bevins2013, Yashiro2015, Pelzer2016}. In contrast, this study presents a fast and simple method using a look-up table (LUT) that enables correction for material combinations of bone and soft tissue in the human thorax. The underlying mathematical principles are illustrated in the following section. 
\subsection{Grating-based X-ray Interferometry}\label{section:Interferrometry}
Grating-based X-ray Interferometry (GBI) employs the Talbot effect to discern USAXS\cite{Pfeiffer2008}, attenuation, and phase\cite{Pfeiffer2006} information using coherent X-ray sources. In this configuration, a phase grating $\text{G}_1$ with a period of $p_1$ generates a Talbot carpet, which exhibits an intensity repetition of the grating period along the beam direction for distances $d_\text{Talbot}(n) = \frac{2p_1^2}{\lambda}\cdot n\cdot k$ and a parallel beam geometry\cite{Suleski1997}. For a cone beam geometry, the distances are adapted to  $d_{\text{Talbot}}^* = \frac{l\cdot d_{\text{Talbot}}}{l-d_{\text{Talbot}}}$\cite{cowley1995}. In this context, $\lambda$ represents the wavelength of the monochromatic X-ray beam, $n$ is a positive integer, and $l$ represents the distance between the focal spot of the coherent X-ray source and $\text{G}_1$. 
The parameter $k$ takes a value of $\frac{1}{4}$ for a  $\frac{\pi}{2}$ phase grating and a value of $\frac{1}{16}$ for a $\pi$ phase grating\cite{Yaroshenko2014} generating the Talbot carpet\cite{Guigay2004, Pfeiffer2005}. Nevertheless, alternatively, an attenuation grating may also be employed as $\text{G}_1$\cite{Huang2009,Olivo2007}. If a further analyzer grating, designated as $\text{G}_2$, is now placed at a position $d_{\text{Talbot}}(n)$, an intensity stepping function can be resolved at the detector for every single pixel. This is achieved by stepping $\text{G}_2$ perpendicular to the beam direction and to the grating lamellae of $\text{G}_1$ and $\text{G}_2$. This function is represented by a first-order Fourier series, which represents an intensity function depending on the $\text{G}_2$ stepping position, and the fit parameters are the amplitude, mean value, and phase \cite{Pfeiffer2008}. Once the fit parameters for amplitude, mean value, and phase for both the sample and reference scans have been determined for an x-ray energy $E$, the attenuation $A(E)$, the dark-field signal $D(E)$, and the differential phase can be derived from these parameters\cite{Noichl2024, Pfeiffer2006}. The phase information is not further discussed in this work. The mathematical representations of $A(E)$ and $D(E)$ useful for this work are expressed by equations \eqref{eq:transmission_monochrom} and \eqref{eq:darkfield_monochrom}. In this context, the variables $\mu$ and $\epsilon$ represent the linear attenuation coefficient and the linear diffusion coefficient, respectively\cite{Bech2010}.
The unit vector pointing from the source to the detector pixel is represented by $\mathbf{\hat{e}_\text{z}}$. The constant $c_D$ depends on the gratings and the geometry of the setup. $V(E)$ is the visibility of the sample scan, $V_0(E)$ is the visibility of the reference scan, $I_0(E) = \psi_{\text{source}}(E) \cdot \mathcal{R}(E)$ is the effective source intensity where $\psi_\text{source}(E)$ ist the intensity from the X-ray tube and $\mathcal{R}(E)$ is the detector response function. $T(E) = \exp\left[-\int_{0}^{z_0} \mu(z\mathbf{\hat{e}_\text{z}},E) dz\right]$ is the damping and $I(E) = I_0(E)\cdot T(E)$ is the intensity detected after passing through a sample or patient\cite{Pfeiffer2007,Bech2010,Lynch2011}.
\begin{align}
&A(E) = -\ln\left(T(E)\right)= -\ln\left(\frac{I(E)}{I_0(E)}\right) = \int_0^{z_0} \mu(z\mathbf{\hat{e}_\text{z}}, E) dz \label{eq:transmission_monochrom}\\[1em]
&D(E) = -\ln\left(\frac{V(E)}{V_0(E)}\right) = -c_D(E) \int_0^{z_0} \epsilon(z\mathbf{\hat{e}_\text{z}}, E) dz \label{eq:darkfield_monochrom}
\end{align}
GBI can also be done using polychromatic non-coherent radiation sources by installing an additional attenuation grating, designated as $\text{G}_0$, behind the source and in front of $\text{G}_1$. This allows for the beam to be modeled coherently as a periodic array of slit sources and enables constructions for clinical use\cite{Pfeiffer2006}.
\subsection{Polychromatic Effects}
For a polychromatic X-ray source, visibility $V_\text{p}$, dark-field signal $D_\text{p}$, and attenuation $A_\text{p}$ are mathematically represented by \eqref{eq:TM_signal}, \eqref{eq:visibility}, and \eqref{eq:DF_signal}\cite{DeMarco2024}. Whereas $V_{\text{0,p}}$ represents the polychromatic visibility of a reference scan at the effective source spectrum $I_0(E)$.
\begin{align}
     &A_\text{p} = -\ln\left(\frac{\int_{0}^{\infty}   T(E)\cdot I_0(E) dE}{\int_{0}^{\infty} I_0(E) dE}\right)\label{eq:TM_signal}\\[1em]
     &V_\text{p} = \frac{\int_{0}^{\infty} V(E) \cdot T(E)\cdot I_0(E)dE}{\int_{0}^{\infty} T(E)\cdot I_0(E)dE} \label{eq:visibility}\\[1em]
     &D_\text{p} = -\ln\left(\frac{V_\text{p}}{V_{\text{0,p}}}\right) \label{eq:DF_signal}
\end{align}
In general, we are interested in the USAXS as additional information for the analysis of matter about the microstructure or, in particular, lung diagnostics because this provides information about the microstructure of the sample\cite{Feigin1987, Yashiro2010, Lynch2011,Strobl2014, Prade2015,Gkoumas2016, Ludwig2019, Graetz2020}. Looking at \eqref{eq:TM_signal}, it can be seen that if attenuation decreases monotonically as a function of energy, the mean energy of the detected spectrum $I(E)$ is monotonically raised while passing through the material, which is known as BH. If we then have a further look at \eqref{eq:visibility}, it is clear that the dark-field signal is influenced by this effect as well, as the hardening of $I(E)$ changes the visibility $V_\text{p}$. Thus, even for a homogeneous material without any microstructure, there is still a change in visibility and, therefore, a dark-field signal\cite{Bevins2013, Yashiro2015, Pelzer2016,DeMarco2024}.\\


\section{Methods}
\subsection{Imaging setup}
\begin{figure}[htb]
    \begin{center}
    \includegraphics[width=0.9\textwidth]{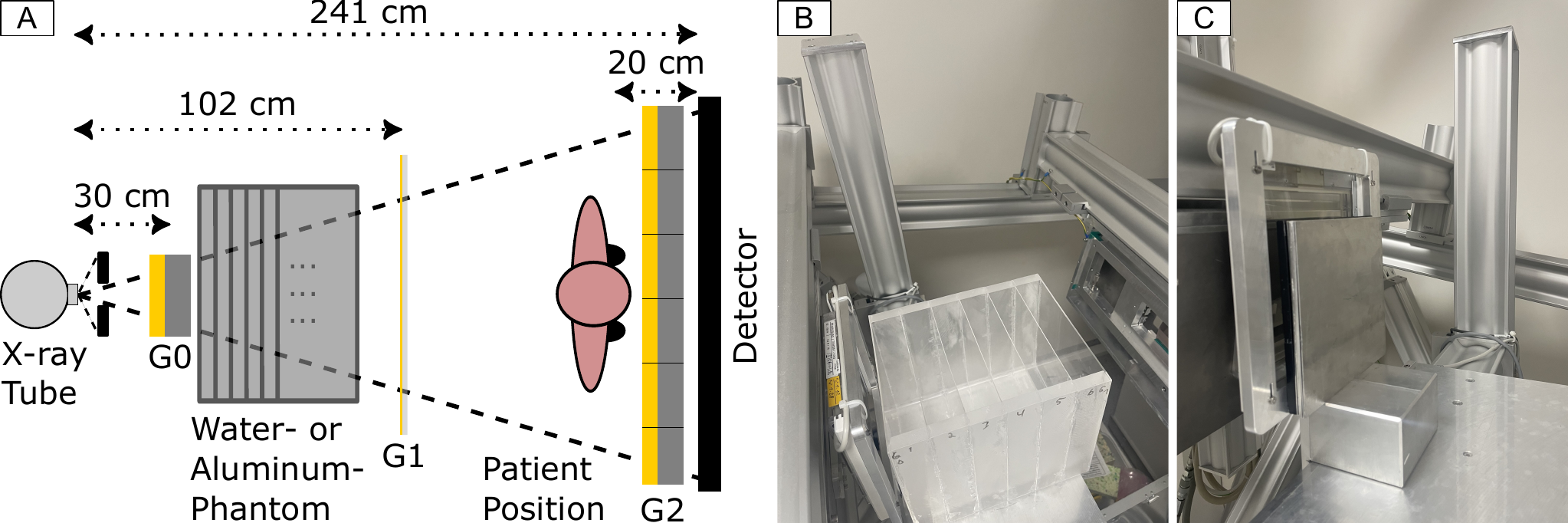}
    \caption{\textbf{A}. The schematic structure of the fringe-scanning system is depicted here from an overhead perspective. The interferometer arm is thus positioned within the drawing plane. The collimated X-ray beam traverses all three gratings. G$_0$ serves to render the polychromatic and incoherent X-ray spectrum coherent. The phase grating G$_1$ generates a Talbot carpet. The stepping function is generated by the interferometer arm movement and a slight mismatch of the grating period of G$_2$ and the period of the intensity pattern of the Talbot carpet at the position of G$_2$, which leads to a moiré pattern at the detector (Tab.\ref{tab:gratingValues}). The thickness of the water phantom or the aluminum phantom was incrementally augmented and positioned between G$_0$ and G$_1$. \textbf{B}. Illustrates the water tank for the soft tissue (behind G$_0$ and before G$_1$) for the calibration. \textbf{C}. Illustrates the aluminum plates for the same purpose and position.}
    \label{figure:phantoms_setup} 
    \end{center}
\end{figure}

The schematic structure of the scanner is illustrated in Fig.\ref{figure:phantoms_setup}A that uses a fringe-scanning scheme\cite{Kottler2007,Arboleda2014}. Principally, this is a classical radiographic system except for the movable interferometer arm placed within the beam path on which three gratings $\text{G}_0$, $\text{G}_1$, and the analyzer grating $\text{G}_2$ are mounted. While $\text{G}_0$ and $\text{G}_1$ are made from single pieces, $\text{G}_2$ is stitched together in six pieces measuring $\SI{6.5}{\centi\meter}$ $\times$ $\SI{7}{\centi\meter}$.\\

Further information about the gratings in terms of fabrication, type, positions, and periods can be seen in Tab.\ref{tab:gratingValues}. Here, a slight deviation in the grating period of $\text{G}_2$ is artificially enforced, which results in a moiré-fringe pattern at the detector. The stepping function itself is then realized by the upward movement of the interferometer arm during the measurement, as the pattern moves over the detector, which is fixed relative to the interferometer. A total of $195$ images are generated during one acquisition of approximately $\SI{7}{\second}$, which are used to determine the stepping function. From that point on, the preliminary dark-field signal and the attenuation channel can be derived as described in \ref{section:Interferrometry} previously.\\
The polychromatic collimated beam is generated by an X-ray tube (MRC 200 0508 ROT-GS 1003, Philips Medical Systems) with filtration of $\SI{2.5}{\milli\meter}$ Al equivalent, the large focal spot of $0.8$, a tube voltage of $\SI{70}{\kilo\volt}$, a pulse rate of $\SI{30}{Hz}$, and a pulse length of $\SI{17.1}{\milli\second}$\cite{Urban2024}.\\
In addition to the tube's intrinsic filtering properties, other components of the setup provide supplementary filtering before reaching the patient plane. These components are the collimator aperture (R 302 MLP/A DHHS, Ralco) located in front of $\text{G}_0$ with $\SI{1}{\milli\meter}$ Al equivalent filtration, the dose-measuring chamber (DIAMENTOR E2, Type 11033, PTW) used to determine the dose-area product (DAP) behind $\text{G}_0$ with $\SI{0.2}{\milli\meter}$ Al equivalent filtration, as well as the filtration of the grids $\text{G}_0$ and $\text{G}_1$ having a combined equivalent filtration of $\SI{7}{\milli\meter}$ Al. The total filtration before the patient, therefore, is $\SI{10.7}{\milli\meter}$ Al equivalence. Subsequently, the beam is subject to further filtration by $\text{G}_2$, which has an equivalence of $\SI{6}{\milli\meter}$ Al. This occurs after the patient and before the detector. The values for the Al equivalent filtrations of the grids were measured during the initial acceptance inspection of the system and were therefore taken from the acceptance report.\\
The detector (PIXIUM 4343 F4, Trixell), having a scintillator layer thickness of $\SI{600}{\micro\meter}$ of CsI, operates with a pixel-binning of 3$\times$3 resulting in a physical pixel size of $\SI{444}{\micro\meter}$. For study participants who are always placed at a contact plane with a distance of $\SI{20}{\centi\meter}$ to the detector, this results in an effective pixel size of about $\SI{400}{\micro\meter}$ in a posterior-anterior (p.-a.) position. This can vary slightly depending on the size of the patient and the distance to the detector. For a reference patient who is scanned in a p.-a. position, the effective dose is $\SI{35}{\micro\sievert}$\cite{Frank2022}.
\newline
\begin{table}[htbp]
\setlength{\tabcolsep}{3pt}
\begin{center}
\begin{tabular} {lcccc}
\hline
&&&&  \vspace{-3mm}\\
      & \textbf{G$_0$}           & \textbf{G$_1$}          &  \textbf{G$_2$}                \\
&&&&  \vspace{-3mm}\\
\hline
\hline
&&&&  \vspace{-3mm}\\
Grating type/influence     & Attenuation           & Phase-Shift          &  Attenuation                \\
&&&&  \vspace{-3mm}\\
Number of tiles         &        1           &        1     &       6     \\
&&&&  \vspace{-3mm}\\
Distance from source ($\SI{}{\centi\meter}$)        &        30           &        102     &       241     \\
&&&&  \vspace{-3mm}\\
Substrate material      & Graphite           & Glass         &  Graphite               \\
&&&&  \vspace{-3mm}\\
Grating material      & Gold           & Gold          &  Gold               \\
&&&&  \vspace{-3mm}\\
Substrate height ($\SI{}{\micro\meter}$)       &    1000           & 200              &    1000              \\
&&&&  \vspace{-3mm}\\
Gold hight ($\SI{}{\micro\meter}$)        &         250      &       9.2        &          250      \\
&&&&  \vspace{-3mm}\\
Grating period ($\SI{}{\micro\meter}$)        &        7.7           &        10.1       &       14.8            \\
&&&&  \vspace{-3mm}\\
\hline
\end{tabular}
\caption{Gratings information of the interferometer for the components G$_0$, G$_1$, and G$_2$.}
\label{tab:gratingValues}
\end{center}
\end{table}

\subsection{Correction of Beam Hardening-induced Dark-field Signal}\label{sec:Correction_BHC}
When considering the the x-ray attenuation by a human thorax, it is appropriate to classify tissues into two main types based on their attenuation characteristics: bone and soft tissue, which do not or only to a very small extent cause real USAXS. Bone and soft tissue both strongly contribute to the hardening of the source spectrum $I_0(E)$, resulting in a BH-induced dark-field signal on top of the real USAXS. However, this component of the dark-field signal does not convey any information about tissue microstructure and should therefore be removed from the dark-field image. Lung tissue, while part of soft tissue, also contributes to the dark-field signal through real USAXS at the alveolar structures. This USAXS-based signal provides valuable insights into the microstructure and represents the component we aim to isolate and visualize in the final dark-field image.\\
Considering equation \eqref{eq:DF_signal}, it can be seen that mathematically, the fraction can be expanded by the BH-induced visibility change caused by an attenuator, so that the expression can be decomposed into a USAXS term $D_{\text{p}}^{\text{'}}$ and a BH-induced term $D_{\text{p}}^{\text{BH}}$ \eqref{eq:BHC_SAXS_separation}. 
\begin{align}
     D_\text{p}
     = -\ln\left(\frac{V_\text{p}}{V_{\text{0,p}}^{\text{BH}}} \cdot \frac{V_{\text{0,p}}^{\text{BH}}}{V_{\text{0,p}}}\right) = -\ln\left(\frac{V_\text{p}}{V_{\text{0,p}}^{\text{BH}}}\right) -\ln\left( \frac{V_{\text{0,p}}^{\text{BH}}}{V_{\text{0,p}}}\right) = D_{\text{p}}^{\text{'}} + D_{\text{p}}^{\text{BH}}
\label{eq:BHC_SAXS_separation}
\end{align}
Here we introduced $V_{\text{0,p}}^{\text{BH}}$ as the reference fringe visibility measured using the same hardened X-ray spectrum as with the sample. Using this reference, we can eliminate the artificial dark-field signal caused by the energy-dependent fringe visibility of the interferometer. The main goal now is to estimate $D_{\text{p}}^{\text{BH}}$ from the actual measurement and some calibration measurements.\\
Here we pursue the concept to use a single LUT to estimate $D_{\text{p}}^{\text{BH}}$. The concept makes use of the fact that the beam hardens monotonically with increasing attenuation. Thus, the measured attenuation $A_{\text{p}}$ can be used as a surrogate for the beam-hardness, i.e., as the entry variable of the LUT.
For a pure attenuator without microstructure, equation \eqref{eq:BHC_SAXS_separation} simplifies to $D_\text{p} = D_\text{p}^\text{BH}$. This allows the BH-induced signal $D_\text{p}^\text{BH}$ and the corresponding attenuation $A_\text{p}$ to be directly measured for various attenuator thicknesses. To calculate the LUTs for the known base materials without any microstructure, they are subjected to individual measurements for different layer thicknesses placed right after $\text{G}_0$ and before $\text{G}_1$ (Fig.\ref {figure:phantoms_setup}). The reason for the different positioning compared to the patient position is that this way, the entire field of view (FOV) can be covered with minimal material, and the exchange of individual layers is easier. Due to very similar attenuation properties, it is common practice in computed tomography (CT) to model soft tissue-induced beam hardening by water \cite{Hsieh2022} and bone by aluminum \cite{Blake1992, Krmar2010}. Therefore, these two materials were used to determine the LUTs for both. In this context, the calibration measurements entail the acquisition of an attenuation image, designated as $A_\text{p}(u,v)$, and a dark-field image $D_\text{p}(u,v)$ for each material and each layer thickness, under the standard measurement protocol. However, given that the measured base materials lack any microstructure and thus do not induce USAXS, any measured dark-field signal is entirely attributable to the hardening of the source spectrum, $I_0(E)$.\\
From the obtained data points, a LUT for pure water $D_{\text{p, H$_2$O}}^{\text{LUT}}(A_{\text{p}}(u,v))$ and for pure aluminum $D_{\text{p, Al}}^{\text{LUT}}(A_{\text{p}}(u,v))$ are created for each pixel $(u,v)$ using spline fits. These were then combined by a weighted mean to a single LUT $D_{\text{p, $\omega_\text{Al}$}}^{\text{LUT}}(A_{\text{p}}(u,v))$ where $\omega_\text{Al} \in[0,1]$ represents the contribution of aluminum \eqref{eq:Weighting_LUTs}. The reason for choosing to perform this correction pixel by pixel is that the intensity spectrum $I_0(E)$ is not globally constant due to the heel effect of the X-ray tube, and the LUT is expected to vary pixel by pixel.
\begin{align}
     D_{\text{p, $\omega_\text{Al}$}}^{\text{LUT}}(A_{\text{p}})  = \left[1 - \omega_{\text{Al}}\right] \cdot D_{\text{p, H$_2$O}}^{\text{LUT}}(A_{\text{p}}) + \omega_{\text{Al}} \cdot D_{\text{p, Al}}^{\text{LUT}}(A_{\text{p}})
\label{eq:Weighting_LUTs}
\end{align}
The final correction procedure for the measurements of the study participants is described by \eqref{eq:BHC_correction_ansatz}. Here, the BH-induced dark-field signal, determined using the LUT and the attenuation image $A_{\text{p}}(u,v)$, is calculated and subtracted from the measured dark-field image $D_{\text{p}}(u,v)$. The term $D_{\text{BHC-bias}}$ is a correction for a bias or overcorrection in regions of pure water or soft tissue that might be present if a  weighted single LUT with $\omega_\text{Al} > 0$ is chosen. Since for X-ray chest radiographs, the exact regions of pure water or soft tissue are unknown and the attenuation changes in the image, this term needs to be chosen for regions of pure water or soft tissue, but is added for all pixels (u,v). Fundamentally, the term $D_{\text{BHC-bias}}$ corresponds to a re-windowing of the image; however, in contrast to conventional windowing, the correction here can be mathematically justified since, in principle, $D_{\text{BHC-bias}}$ can be calculated for a region of pure water by the difference between the LUT of pure water and the weighted single LUT used \eqref{eq:bias_formula}.
\begin{align}
     &D_{\text{p}}^{\text{'}}(u,v) = D_{\text{p}}(u,v) - D_{\text{p, $\omega_\text{Al}$}}^{\text{LUT}}(A_{\text{p}}(u,v)) + D_{\text{BHC-bias}} \quad \forall (u, v)\label{eq:BHC_correction_ansatz}\\[1em]
     &D_{\text{BHC-bias}}(A_\text{p, H$_2$O}) = \omega_{\text{Al}}\cdot \left[D_{\text{p, H$_2$O}}^{\text{LUT}}(A_\text{p, H$_2$O}) - D_{\text{p, Al}}^{\text{LUT}}(A_\text{p, H$_2$O})\right]\label{eq:bias_formula}
\end{align}
Thus, the goal of this work is to find optimal value pairs of  $\omega_\text{Al}$ and $D_{\text{BHC-bias}}$ to achieve the best reduction of the BH-induced dark-field signal and the bony structures in the final images. Therefore, study participant measurements were processed with different settings of this beam hardening correction (BHC) method and analyzed.
\subsection{Phantom Constructions}
For the base materials, total thicknesses of $\SI{24.0}{\centi\meter}$ for water and $\SI{6.0}{\centi\meter}$ for aluminum were chosen. For the aluminum phantom, several aluminum sheets with thicknesses of $\SI{0.2}{\centi\meter}$ and $\SI{0.1}{\centi\meter}$ were used. In total, 13 different thicknesses were realized. \\
In the case of the water phantom, a corresponding water container was made of Poly(methyl 2-methylpropenoate) (PMMA), which has six separate chambers, each with a thickness of $\SI{4}{\centi\meter}$. This means that the chambers can be gradually filled with water from measurement to measurement, allowing measurements to be made for a total of seven different water thicknesses. The wall thickness of the PMMA was $\SI{0.1}{\centi\meter}$ for the inner partitions and $\SI{0.15}{\centi\meter}$ for the end walls. All individual material thicknesses of aluminum and water are summarized in Tab.\ref{tab:calibration_thicknesses}. Photos of the phantoms are shown in Fig.\ref{figure:phantoms_setup}B,C.\\

\begin{table}[htbp]
\setlength{\tabcolsep}{3pt}
\begin{center}
\begin{tabular}{lccccccccccccc}
\hline
&&&&  \vspace{-3mm}\\
Thickness Number & 1 & 2 & 3 & 4 & 5 & 6 & 7 & 8 & 9 & 10& 11& 12& 13\\
&&&&  \vspace{-3mm}\\
\hline
\hline
&&&&  \vspace{-3mm}\\
Water ($\SI{}{\centi\meter}$)     & 0$^*$ & 0 & 4 & 8 & 12 & 16 & 20 & 24 & - & -& -& -& -\\
&&&&  \vspace{-3mm}\\
Aluminum ($\SI{}{\centi\meter}$)         & 0.0 & 0.2 & 0.4 & 0.6 & 0.7 & 1.1 & 1.4 & 2.1 & 2.5 & 3.0 &3.9 & 5.0 & 6.0 \\
&&&&  \vspace{-3mm}\\
\hline
\end{tabular}
\caption{Material thicknesses of the water- and aluminum-phantoms. *Empty image (measured without the water container).}
\label{tab:calibration_thicknesses}
\end{center}
\end{table}
\subsection{Application on Study Participants}
In order to ascertain the optimal aluminum weighting $\omega_\text{Al}$ of the single LUT \eqref{eq:Weighting_LUTs} and the level of the BHC-bias correction $D_\text{BHC-bias}$, three dark-field radiographs in p.-a. orientation were corrected by the BHC method presented in \ref{sec:Correction_BHC} for different combinations of $\omega_\text{Al}$ and $D_\text{BHC-bias}$. The goal is to find the best reduction of the bony structures and BH-induced dark-field signal - the best improvement of image quality. For this purpose, analyses were conducted on both the lung regions and the extrapulmonary areas near the clavicle, where no real USAXS signal is anticipated due to the lack of microstructural features.\\
The larger prospective study out of which the three study participants for this work were taken was conducted in accordance with the Declaration of Helsinki (as revised in 2013). Approval of the institutional review board and the national radiation protection agency was obtained prior to this study (Ethics Commission of the Medical Faculty, Technical University of Munich, Germany; reference no. 587/16S). Participants gave their written informed consent.\\
To ensure the target detector dose of $\SI{3.75}{\micro\gray}$ was achieved, the required tube current for each study participant was individually determined based on a calibration curve and dose reference data obtained from the conventional imaging system.\cite{Lochschmidt2025}.\\
Care was taken to include one participant with a heavy physique and two with medium or light physiques in order to cover the typical range of X-ray attenuation observed in the human body. These participants were also chosen to include one healthy participant and two participants with chronic obstructive lung disease (COPD) to ensure that the typical range of dark-field signal in humans was also represented. The gender, height, and weight are given in Tab.\ref{tab:patients_data}. For participants with COPD, it could be shown previously that the fringe scanning prototype provided dark-field images that improved the diagnostics \cite{Willer2021, Urban2022emph, Urban2023emph}.\\

\begin{table}[htbp]
\setlength{\tabcolsep}{3pt}
\begin{center}
\begin{tabular} {lcccc}
\hline
&&&&  \vspace{-3mm}\\
      & \textbf{Healthy}           & \textbf{COPD 1}          &  \textbf{COPD 2}                \\
&&&&  \vspace{-3mm}\\
\hline
\hline
&&&&  \vspace{-3mm}\\
Gender    & Male           & Male          &  Male                \\
&&&&  \vspace{-3mm}\\
Weight ($\SI{}{\kilo\gram}$)         &        96.2           &        64.1     &       64.3     \\
&&&&  \vspace{-3mm}\\
Height ($\SI{}{\centi\meter}$)      &        170           &        173     &       175     \\
&&&&  \vspace{-3mm}\\

\hline
\end{tabular}
\caption{Gender, weight, and height of one healthy and two COPD study participants.}
\label{tab:patients_data}
\end{center}
\end{table}
\subsection{Additional Image Pre-corrections}
For study participants, additional standard corrections were implemented for all measurements conducted on the setup before the BHC presented here was applied. Firstly, Compton scatter caused by the gratings in the beam path, which is dominated by $\text{G}_2$, are considered. This correction is based on a scatter kernel, which was defined using simulations using the Geant4 toolkit (version 10.06.p03)\cite{Geant4,Urban2024}. Compton scattering caused by study participants is corrected by SkyFlow (Philips Medical Systems \cite{skyflow}), adapted to the scanning system. Finally, the detector itself has a non-negligible detector crosstalk. A specific kernel was also defined for this purpose, which removes this crosstalk\cite{Urban2024}. Moreover, motion corrections for the heart, vibration corrections resulting from interferometer arm movement, and bias correction are applied to the images\cite{koehler2023bias,Schick2022,Noichl2024}.\\
The Compton scatter correction using SkyFlow had to be manually turned off for the water and aluminum phantoms due to the changed positioning compared to the study participants. Instead, the sample scatter for each layer thickness was simulated with the Geant4 toolkit. The reason for tuning off SkyFlow is that the software is only adapted for positioning objects at the patient position $\SI{20}{\centi\meter}$ before the detector. For varying distances, the scatter must be manually simulated. As was the case for the study participants, all other corrections remained activated.
      

\section{Results}
\subsection{Look-up Tables for Beam Hardening Correction}
\begin{figure}[t]
    \begin{center}
    \includegraphics[width=0.9\textwidth]{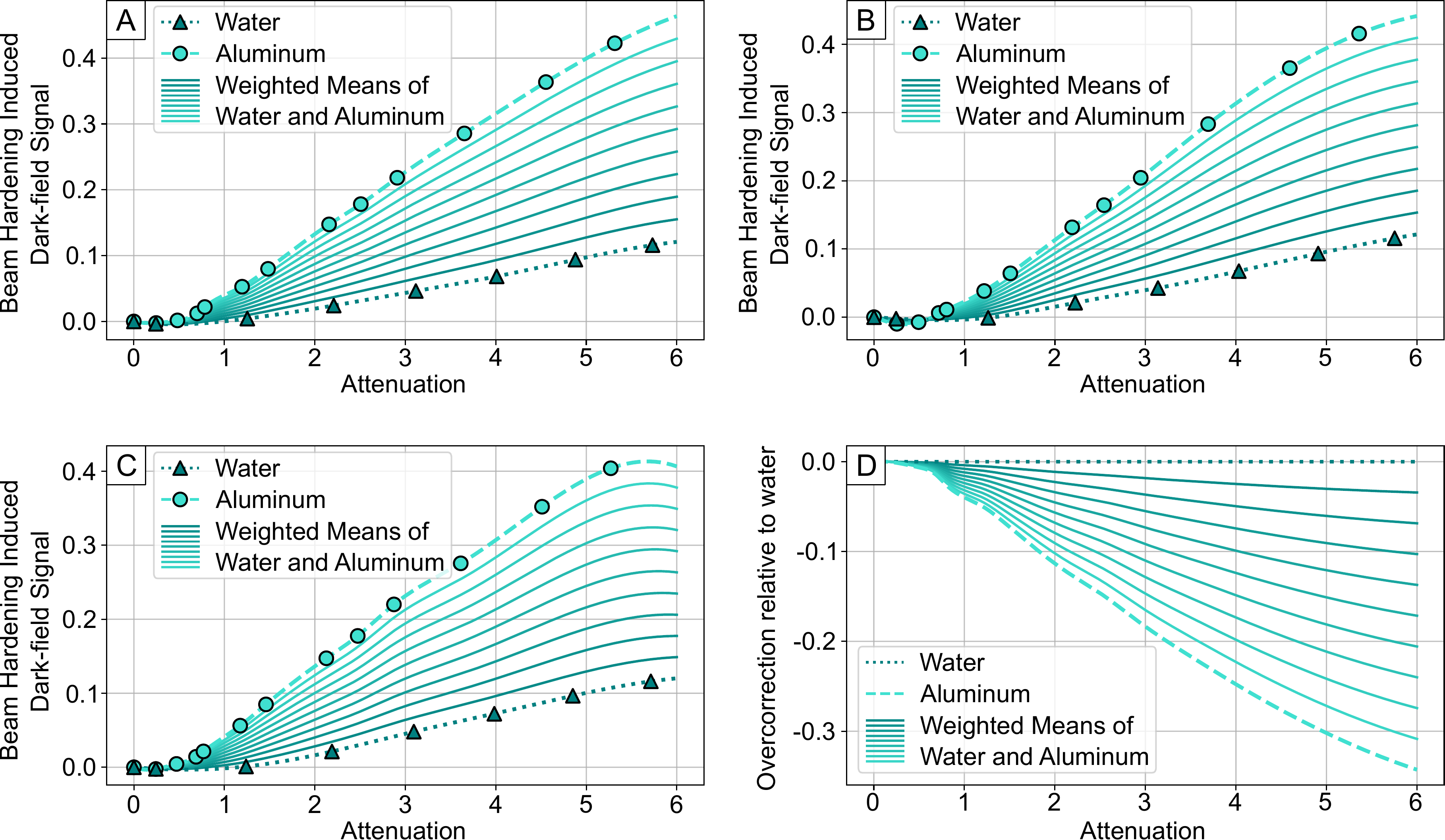}
    \caption{\textbf{A}. Illustration of the dark-field signal induced by BH for the aluminum and water container phantoms (Tab.\ref{tab:calibration_thicknesses}), plotted over the corresponding attenuation for the middle, \textbf{B}. for the upper and \textbf{C}. for the bottom region of the detector. The LUTs for aluminum and water, generated by a spline-function fit, and different weighted averages of both are displayed. \textbf{D}. Shows the overcorrection compared to pure water regions for different aluminum weightings $\omega_\text{Al}$ and is mathematically described by \eqref{eq:bias_formula}.}
    \label{figure:setup_and_splineplots} 
    \end{center}
\end{figure}

As previously described, the spline-fit functions or LUTs are created pixel by pixel. Here, a representative region of interest (ROI) was selected from the center, top, and bottom regions of the detector to show the spline fits or LUTs for pure water, pure aluminum, and weighted single LUTs in increments of 10\% of $\omega_\text{Al}$ as described by \eqref{eq:Weighting_LUTs} and shown in Fig.\ref{figure:setup_and_splineplots}A-C. For the center region, the BHC-bias or overcorrection relative to pure water is shown in Fig.\ref{figure:setup_and_splineplots}D. All single LUTs and the plotted BHC-bias serve as the basis for the various corrections to the study participants.\\
\subsection{Application on Study Participants}
\begin{figure}[htbp]
    \begin{center}
    \includegraphics[width=0.9\textwidth]{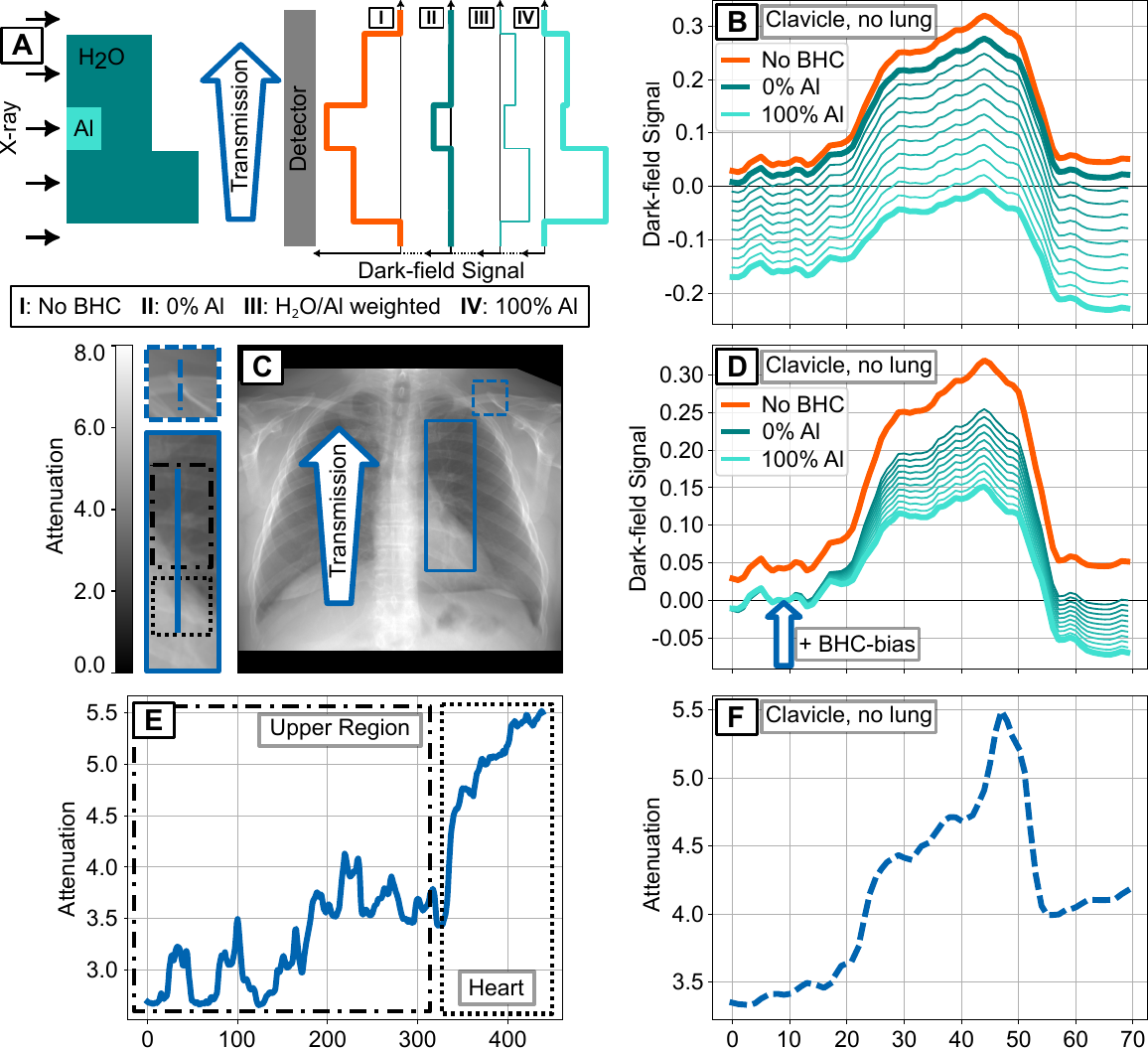}
    \caption{\textbf{A}. Illustration of the changing bone level to pure water with different weightings of the aluminum and water LUT to a single LUT and different transmission levels. \textbf{B}. Lineplots of the BH-induced dark-field signal and \textbf{F}. The attenuation signal through the left clavicle for different LUT weightings and for the healthy patient with a heavy physique. \textbf{D}. represents a correction of the BHC-bias for the lower attenuation region. \textbf{E}. Line plot of the range of the attenuation from the upper part of the lung down to the region of the heart. \textbf{C}. The attenuation image of the healthy patient shows where the line plots were taken.}
    \label{figure:concept_LUT_weightings} 
    \end{center}
\end{figure}

\begin{figure}[htbp]
    \begin{center}
    \includegraphics[width=0.9\textwidth]{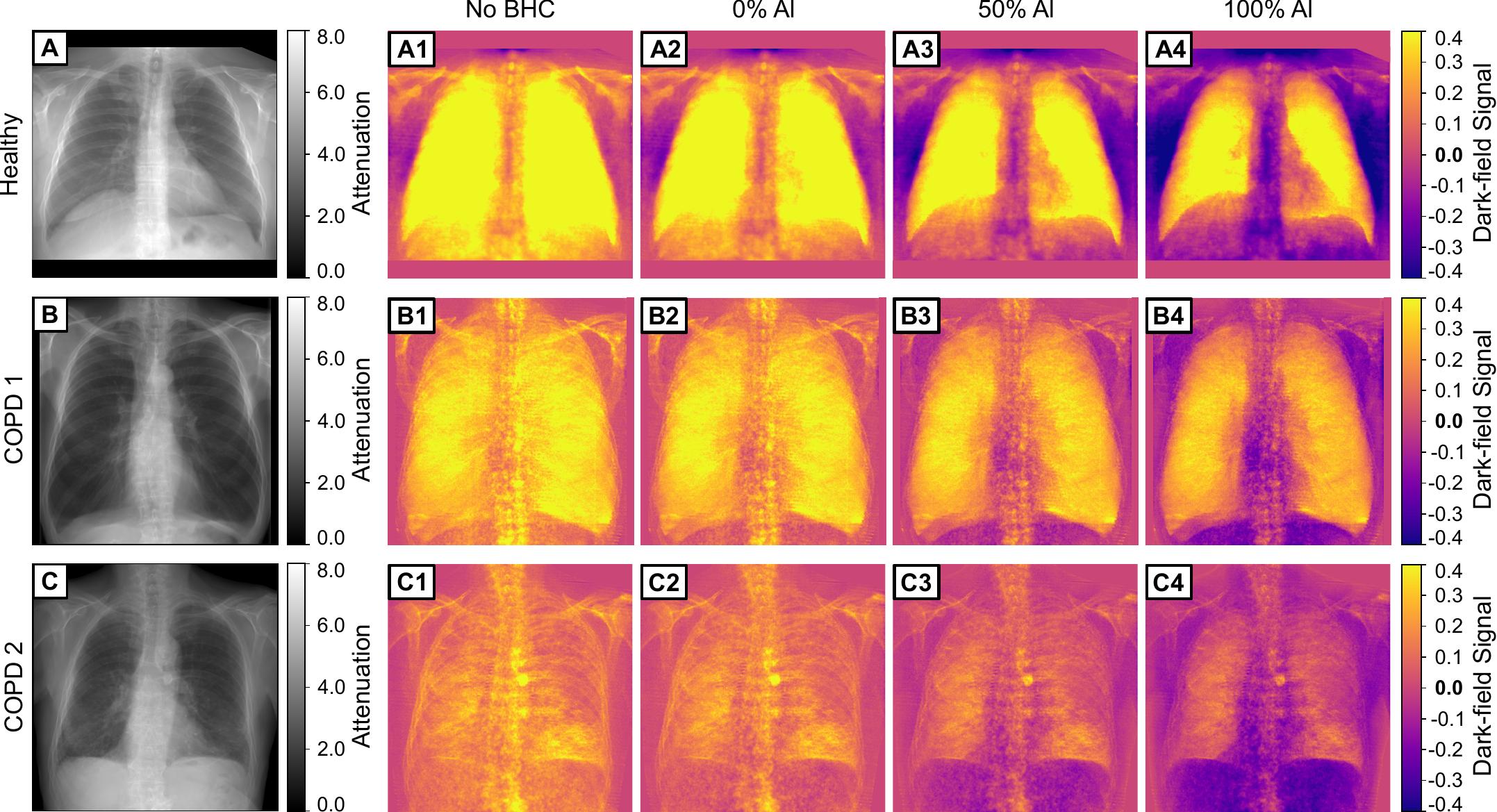}
    \caption{\textbf{A-C}. Attenuation images of the healthy and the two COPD patients. Next to them, the effect of the single LUT corrections on the images regarding the overcorrection described in more detail in Fig.\ref{figure:concept_LUT_weightings} is illustrated for the LUT weightings of \textbf{A2,B2,C2}. 0\% aluminum, \textbf{A3, B3, C3}. 50\% aluminum, \textbf{A4,B4,C4}. 100\% aluminum, and \textbf{A1, B1, C1}. without any BHC correction.} 
    \label{fig:overview_overcorrection} 
    \end{center}
\end{figure}

\begin{figure}[htb]
    \begin{center}
    \includegraphics[width=0.9\textwidth]{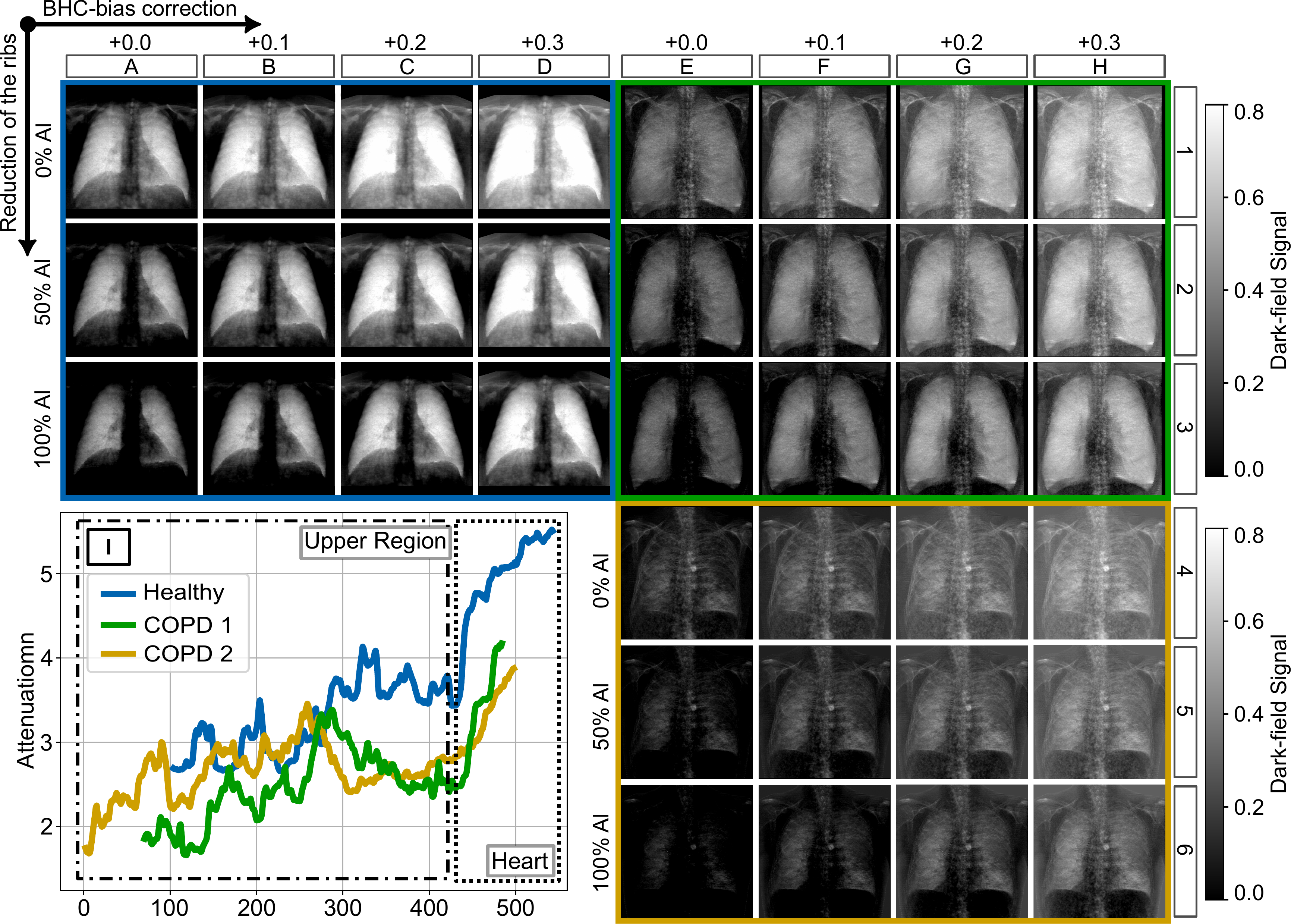}
    \caption{\textbf{A1-H6}. Illustrates the healthy patient (blue) and the two COPD patients (green and yellow) being processed for different bias correction levels and for the LUT weightings of 0\%, 50\%, and 100\% of aluminum. \textbf{I}. Represents line plots of the attenuation images of all patients from the upper part of the lung to the region of the heart.}
    \label{comparison_bias_omega} 
    \end{center}
\end{figure}

In an initial analysis, the general impact of this BHC and the selection of the weighting parameter $\omega_\text{Al}$ on the occurrence of false USAXS signals was investigated using data from the healthy study participant. A line profile across the clavicle bone, located outside the lung region (Fig.\ref{figure:concept_LUT_weightings}B), clearly demonstrates that increasing the aluminum weighting $\omega_\text{Al}$ leads to a reduction in the bone step, thereby enhancing signal homogeneity in this region.\\
However, this improvement is accompanied by a systematic increase in BHC-bias, with the signal being progressively shifted toward negative values as $\omega_\text{Al}$ increases. This observation confirms that the magnitude of the BHC-bias is not solely dependent on the weighting factor but is also influenced by the local attenuation level, as also illustrated in Fig.\ref{figure:setup_and_splineplots}D.
This dependency is particularly evident in Fig.\ref{figure:concept_LUT_weightings}D, where the BHC-bias parameter $D_\text{BHC-bias}(A_\text{p})$ was individually adjusted for each aluminum weighting such that the attenuation region to the left of the bone step was corrected to zero, effectively eliminating the overcorrection. Nevertheless, a residual overcorrection remains on the right side of the bone step, corresponding to the differential attenuation between the two regions. This phenomenon is further exemplified in Fig.\ref{figure:concept_LUT_weightings}A.
The clinical relevance of attenuation-dependent overcorrection becomes even more apparent when examining a line profile extending from the upper lung region to the cardiac region (Fig.\ref{figure:concept_LUT_weightings}C,E). A gradual increase in attenuation is observed along this path. Consequently, an exact bias correction term $D_\text{BHC-bias}$ strongly depends on attenuation for clinical radiographic dark-field images.\\
The effects of aluminum weighting and attenuation on the overall dark-field image are presented in Fig.\ref{fig:overview_overcorrection} for all three study participants. As in previous observations, it becomes particularly evident—especially in regions outside the lung—that higher attenuation signals, at constant aluminum weighting, generally lead to an increased BHC-bias. Conversely, for a given attenuation level, an increase in aluminum weighting also results in a higher BHC-bias and a reduction of the bone steps.\\
Overall, it can be seen that the choice of $\omega_\text{Al}$ governs the degree of the bone step reduction, while the BHC-bias parameter $D_\text{BHC-bias}$ compensates for overcorrection within a targeted attenuation range. Both parameters are compared across all three patients in Fig.\ref{comparison_bias_omega}, alongside the corresponding attenuation profiles extending from the upper lung region to the cardiac region.\\

\section{Discussion} 
It should be noted that the method presented in this work, as described by equations \eqref{eq:BHC_SAXS_separation}, \eqref{eq:Weighting_LUTs}, \eqref{eq:BHC_correction_ansatz}, and \eqref{eq:bias_formula}, is a pixel-wise correction of BH–induced dark-field signals, which needs parameters chosen before, since the material composition of bone and soft tissue is not known for every pixel. However, the method has the ability to significantly reduce both the fake BH signal and bone structures in the dark-field image for regions with similar attenuation, provided that the parameters $\omega_\text{Al}$ and $D_\text{BHC-bias}$ are appropriately chosen. This effect is particularly evident in the overview of all three study participants shown in Fig.\ref{comparison_bias_omega}.
Furthermore, it becomes clear that there is no single pair of values for both parameters that consistently yields the best image correction. For each study participant, the attenuation in the lung region must be assessed in order to determine the appropriate bias correction term $D_\text{BHC-bias}$ using the plot shown in Fig.\ref{figure:setup_and_splineplots}D, and based on the selected aluminum fraction $\omega_\text{Al}$. Depending on the clinical question, these parameters have to be chosen, or compromises in both parameters may be acceptable to achieve a global rather than region-specific correction of the dark-field image.\\
One reason this method was applied on a pixel-wise basis is the presence of the heel effect. Due to the mounting configuration of the X-ray tube, the heel effect is expected to occur vertically along the detector in such a way that the mean energy of the source spectrum $I_0(E)$ becomes progressively higher from the upper to the lower part of the detector. This can also be observed in Fig.\ref{figure:setup_and_splineplots}A–C as a horizontal shift in the measurement points used to determine the LUTs for pure water and pure aluminum. However, this effect is minor, and therefore only the central region of the detector was considered for the overcorrection plot in Fig.\ref{figure:setup_and_splineplots}D and for further usage of the bias correction term $D_\text{BHC-bias}$.\\
It is also important to discuss to what extent other established methods can be applied to clinical radiographic lung imaging. If an object consisted only of soft tissue, the problem could be solved by simulations or a calibration measurement using a soft-tissue-like material\cite{Pelzer2016}. However, this does not work for the two-material case at hand. Another way is to define a single calibration curve for several single materials by weighting the measured transmission with the calculated transmission at the design energy of the imaging system\cite{Yashiro2015}. However, given the dependence of these weighting factors on the material, the correct weighting factor for material  combinations or mixtures remains unknown. Consequently, the applicability of this method to clinical radiographic lung images is also precluded. One further simple method is a stronger filtering of the source spectrum $I_0(E)$ \cite{Bevins2013}. However, this can only slightly mitigate the effect of BH and reduces visibility.\\
Whereas, the method presented in this work represents an improvement for imaging scenarios involving combinations of bone and soft tissue, particularly in clinical radiographic lung imaging. At the same time, it encourages further research in this direction. One improvement might be to use the attenuation image for bone segmentation, resulting in bone- and soft tissue-masks. For each mask, different weightings of the LUTs could then be chosen and applied to the dark-field image being corrected for the fake dark-field signal. In this case, as well, a weighting would need to be estimated for the regions identified as containing bone. However, such an approach would eliminate the need for a separate bias correction, since it would be known which regions are affected by soft tissue only and which involve a combination of materials, including bone. A further improvement might be a pixel-wise, physically exact correction, without any discussion about the LUT weightings, which is only possible if attenuation can be separated for each material. Using spectral radiographic measurements, the artificial dark-field signal could then be corrected using a model of the dark-field and attenuation signals, considering all materials involved for each pixel.\\


\section{Conclusions}
In conclusion, we proposed a BHC using a single LUT to reduce the artificial BH signal and the bony structures for dark-field chest imaging to a large extent. The approach is LUT-based semi-empirical, as the content of the LUT is based on physical measurements, but the specific weighting to generate the final LUT values is optimized using qualitative image quality assessment. Practically, two parameters and their respective effects on the corrected images must be considered when applying this type of correction. First, the aluminum weighting, which determines the extent to which rib structures are suppressed, and second, the BHC bias correction term compensating for the overcorrection introduced by this method. The appropriate level of BHC bias correction is strongly dependent on the local attenuation and can optionally be done by re-windowing of the dark-field images as well.
Therefore, for each clinical question, the parameter configuration must be individually assessed to ensure sufficiently high image quality, or compromises in both parameters may be acceptable to achieve a global rather than region-specific correction of the dark-field image. In principle, however, an optimal configuration can always be found for each region of the lung, enabling effective correction of the BH-induced dark-field signal and bony structures.
\section{Acknowledgments}
This work was supported by the European Research Council (ERC, H2020, AdG 695045), the Deutsche Forschungsgemeinschaft (GRK 2274), the Federal Ministry of Education and Research (BMBF) and the Free State of Bavaria under the Excellence Strategy of the Federal Government and the Länder, the Technical University of Munich – Institute for Advanced Study, and Philips Medical Systems DMC GmbH. This work was carried out with the support of the Karlsruhe Nano Micro Facility (KNMF, www.kit.edu/knmf), a Helmholtz Research Infrastructure at Karlsruhe Institute of Technology (KIT).

\section{Conflict of Interest Statement}
The authors declare no competing interests.



\section*{References}

\addcontentsline{toc}{section}{\numberline{}References}
\vspace*{-12mm}










\bibliographystyle{./medphy.bst} 

\end{document}